\documentclass[preprint]{revtex4}

\usepackage{graphicx}

\newcommand{\beq}{\begin{equation}}
\newcommand{\enq}{\end{equation}}
\newcommand{\bea}{\begin{eqnarray}}
\newcommand{\ena}{\end{eqnarray}}
\newcommand{\rr}{{\bf r}}
\newcommand{\RR}{{\bf R}}
\newcommand{\aos}{a_{\mathrm{osc}}}

\begin{document}
\title{Rotational states of Bose gases with attractive interactions
in anharmonic traps}
\author{Emil Lundh}
\altaffiliation[Present address: ]{Department of Physics, 
KTH, SE-10691 Stockholm}
\affiliation{Helsinki Institute of Physics, PL 64, FIN-00014
Helsingin yliopisto, Finland}
\author{Anssi Collin}
\affiliation{Helsinki Institute of Physics, PL 64, FIN-00014
Helsingin yliopisto, Finland}
\author{Kalle-Antti Suominen}
\affiliation{Helsinki Institute of Physics, PL 64, FIN-00014
Helsingin yliopisto, Finland}
\affiliation{Department of Physics, University of Turku, FIN-20014
Turun yliopisto, Finland}
\date{\today}

\begin{abstract}
A rotated and harmonically trapped Bose gas with attractive interactions is 
expected to either remain stationary or escape from the trap. 
Here we report that, on the contrary, 
in an anharmonic trapping potential the Bose 
gas with attractive interactions 
responds to external rotation very 
differently, namely through center-of-mass motion or by formation of vortices, 
depending on the strengths of the interaction and the anharmonic term in 
the trapping potential. 
\end{abstract}


\maketitle

Trapped ultracold gases of atoms with attractive {\it s}-wave interactions offer quite different physics compared to the more common case of repulsive interactions \cite{pethicksmith}. The behavior under rotation is an important example. 
While repulsive Bose-Einstein condensed gases respond to rotation by forming quantized vortices, such states are not necessarily energetically favorable in attracting gases. In fact, Wilkin {\it et al.}~\cite{wilkin} have studied a {\it harmonically} trapped gas of attracting bosons under rotation in the limit where the interactions can be treated perturbatively. They find that in the lowest-energy many-body state for a given angular momentum (the yrast state) the center of mass (CM) carries all the angular momentum. According to Wilkin {\it et al.}, for this state the single-particle density matrix has more than one eigenvalue of macroscopic size, which means that more than one single-particle state is macroscopically occupied. Such a state is called a fragmented condensate~\cite{nozieres}. The form of the lowest-energy state has since been proved in greater generality~\cite{hussein}. 
Pethick and Pitaevskii~\cite{pethickpitaevskii} made the important observation that the rotational state in question can be seen as a Bose-Einstein condensate if it is viewed in the CM reference frame. 
Nevertheless, it seems that this interesting state of motion is never thermodynamically stable, because the critical frequency is equal to the frequency for destabilization of the cloud. On the contrary, in this Letter we show that an {\it anharmonic} trapping potential makes the state of CM rotation, as well as vortex states, attainable.

We consider a trapped, attractively interacting Bose gas at zero temperature, subject to a rotational force with a given angular velocity, and study the states that minimize the energy under such conditions. 
To this end, we first derive an equation of motion governing such a gas, taking into account the CM motion. 
We then study numerically and variationally the behavior in an anharmonic trapping potential, and show that the parameter space is divided into regions of CM motion and vortex formation.

The system under study consists of $N$ bosons subject to the many-body Hamiltonian 
\beq\label{hamiltonian}
   H = -\frac{\hbar^2}{2m}\sum_{i=1}^N\frac{d^2}{d\rr_i^2} 
   +\sum_{i=1}^N V(\rr_i)
   + \frac{U_0}{2}\sum_{i\neq j} \delta(\rr_i-\rr_j).
\enq
The external potential $V(\rr)$ can in experiment be of either magnetic or optical origin, and is usually taken to be a harmonic-oscillator well. But the harmonic potential has special properties that simplify the problem but in fact prevent rotation, and we therefore study the more general case of a cylindrically symmetric trap that is anharmonic in the radial direction:
\beq\label{anharmpotential}
   V(r,\theta,z)=\frac12m\left[\omega^2 (r^2 + \lambda \frac{r^4}{\aos^2})+ 
   \omega_z^2 z^2\right].
\enq
The oscillator length $\aos$ is defined as $\aos=(\hbar/m\omega)^{1/2}$. The third term in the Hamiltonian (\ref{hamiltonian}) is the inter-atomic interaction potential which is here replaced by its usual {\it s}-wave approximation; the strength $U_0=4\pi\hbar^2a/m$ depends on the 
scattering length $a$ which in this study will be assumed negative, implying an attractive interaction.

For the case of purely harmonic trapping ($\lambda=0$), the state of CM rotation was discussed in Refs.~\cite{wilkin,hussein,mottelson}. The energy of $L$ quanta of angular momentum carried by the CM is exactly $L\hbar\omega$, independent of interactions~(\cite{fetterrokhsar}, Appendix A; cf.~\cite{kohn,elliottskyrme}). This has the important consequence that the gas cannot be rotated, because the critical frequency for the excitation of any rotation is equal to the frequency where the cloud is destabilized. The energy of a system in CM rotation as seen in a frame rotating with angular velocity $\Omega$ is, namely,
\beq\label{harmenergy}
   E -\Omega L_z = E_0 + \hbar (\omega - \Omega) L,
\enq
where $E_0$ is the energy of the cloud in the absence of rotation. The energy, Eq.~(\ref{harmenergy}), is minimized for $L=0$ when $\Omega<\omega$ and for 
$L\rightarrow \infty$ when $\Omega > \omega$. Consequently, a harmonically trapped attracting Bose gas cannot be set into steady rotational motion: it either stays still, for slower 
stirring, or escapes from the trap for faster stirring. 

In order to stabilize a finite-angular momentum state one may add an anharmonic term to the trapping potential. In such a potential, the single-particle energy as a function of angular momentum increases faster than linear, 
and the energy in a rotating frame may therefore have a minimum for a finite value of $L$. However, there is {\it a priori} no guarantee that the lowest-energy state is that of CM rotation when the trap is anharmonic. In fact, for a noninteracting gas in an anharmonic trap it is known that the 
lowest-energy state for given angular momentum $L_z=Nm_l\hbar$ is the state where all particles are in the same angular momentum eigenstate, in other words, a Bose-Einstein condensate containing a vortex with quantum number $m_l$~\cite{lundh2002}. This is so because the nonlinear dependence of the energy on angular momentum implies that an angular-momentum eigenstate has lower energy than any superposition of different states with the same expectation value of angular momentum. 
An intuitive picture can be constructed as follows: in a frame rotating with a fixed angular velocity $\Omega$, the sum of the centrifugal and potential energies create a Mexican-hat shaped potential if the angular velocity is high enough. In the absence of interactions, the lowest-energy state is one of toroidal shape lying along the minimum of the effective potential. When the interactions are attractive, however, the tendency for the density to be maximized may favor the state of CM rotation.
It remains to investigate the crossover between these two types of rotational state.

We now put up an ansatz that allows for the derivation of a simple set of equations of motion of an attractive gas. In a harmonically confined gas with isotropic interactions, the CM and relative motions decouple and therefore the wave function of the $N$-body system can be written in the form
\beq\label{mbwf1}
   \Psi(\rr_1,\dots,\rr_N) = \psi_C(\RR)\psi_R(\rr'_1,\ldots,\rr'_N),
\enq
where $\RR= (\rr_1+\ldots+\rr_N)/N$ is the CM coordinate and 
$\rr'_i=\rr_i-\RR$ are the coordinates of each particle relative to the 
CM, $i=1,\ldots,N$~\cite{pethickpitaevskii,elliottskyrme}. Furthermore, 
the relative part $\psi_R$ exhibits Bose-Einstein condensation, so that 
the associated density matrix has one macroscopic eigenvalue~\cite{pethickpitaevskii}. 
Consequently, the many-body wave function can be written on the form
\beq\label{trialwf}
   \Psi(\rr_1,\ldots,\rr_N) = \psi_C(\RR) \prod_{i=1}^N \chi(\rr'_i),
\enq
as long as the depletion of the condensate 
is neglected. Although Eqs.~(\ref{mbwf1}-\ref{trialwf}) hold strictly only in the case of a harmonic trap, we shall assume that the decoupling between CM and internal motion holds approximately also in the anharmonic case and use Eq.~(\ref{trialwf}) as an ansatz for the derivation of an equation of motion that can be used to study the rotational properties 
of the attractive gas. We expect the resulting equation to be approximately 
valid when $\lambda$ is small so that the trap does not deviate much from harmonic.

The interaction energy is independent of the center-of-mass motion, simply because
\beq\label{distance}
   \rr_i-\rr_j = \rr'_i-\rr'_j.
\enq
The harmonic part of the trap potential energy can be expressed in the relative coordinates according to
\beq\label{harmonic}
   \sum_i {x}_i^2 = N{X}^2 + \sum_i {x'_i}^2,
\enq
and equivalently for the other Cartesian components. The results 
(\ref{distance}-\ref{harmonic}) together imply that in the harmonic case, 
the CM motion is decoupled from the internal motion, as discussed above. 
The quartic part of the potential expressed in CM and relative coordinates 
is
\bea
   \sum_{i=1}^{N} r_i^4 = NR^4 +\sum_{i=1}^{N}\left[
   2R^2{r'_i}^2 + 4(Xx'_i+Yy'_i)^2 + 
   4{r'_i}^2(Xx_i'+Yy'_i) + {r'_i}^4 \right].
\label{quartictransfo}
\ena

The equations of motion are to be obtained in a variational procedure, slightly generalizing the derivation given in 
Refs.~\cite{pethicksmith,leggett}. 
We express the energy of the $N$-body system in terms of 
the trial wave function, Eq.~(\ref{trialwf}),
\beq
   E = \int d\rr_1\ldots d\rr_N \Psi^*(\rr_1,\ldots,\rr_N) H 
   \Psi(\rr_1,\ldots,\rr_N),
\enq
where $H$ is the Hamiltonian, Eq.~(\ref{hamiltonian}). The transformation to the new system of $N+1$ coordinates, $\RR,\rr'_1,\ldots,\rr'_N$, introduces a Jacobian $J=N^d$ in $d$ 
dimensions and 
a delta function $\delta(\rr'_1+\rr'_2+\ldots+\rr'_N)$ in the integral, because the relative coordinates are not 
independent. Neglecting this interdependence of the coordinates introduces only errors of order $N^{-1}$, however, so that the $N+1$ variables can in effect be regarded as independent and the functions $\psi_C$ and $\chi$ can be assumed to be individually normalized to unity. The trap potential term is rewritten according to Eqs.~(\ref{harmonic}-\ref{quartictransfo}), and 
the action of the kinetic-energy operator on the trial wave function is readily calculated. 
With a number of superfluous factors $\chi(\rr'_i)$ integrated out, the energy functional is
\bea
   E = \int J d\RR d\rr' \psi_C^*(\RR) \chi^*(\rr') 
   \left[ -\frac{\hbar^2}{2Nm}\frac{\partial^2}{\partial \RR^2}
   + \frac{Nm}{2}(\omega^2 R^2 + \omega_z^2 Z^2)+\right.\nonumber\\ 
   N\lambda \frac{m\omega^2}{2\aos^2}
   \left(R^4+2R^2{r'}^2
   +4(Xx'+Yy')^2+4{r'}^2(Xx'+Yy')+ {r'}^4\right)\nonumber\\
   \left.
   -(N-1)\frac{\hbar^2}{2m}
   \frac{\partial^2}{\partial \rr^2} 
   + N\frac{m}{2} (\omega^2 {r'}^2  +\omega_z^2 {z'}^2 )
   + N(N-1) \frac{U_0}{2} |\chi(\rr')|^2\right]
   \psi_C(\RR) \chi(\rr'),
\label{energy}
\ena
where $J=N^3$ as noted above. The energy is now functionally minimized with respect to the two wave functions $\psi_C$ and $\chi$. The result is the two coupled equations
\bea
   \left\{-\frac{\hbar^2}{2M}\frac{\partial^2}{\partial \RR^2} 
   + V_C(\RR) \right\}\psi_C(\RR) = E_C \psi_C(\RR),
   \label{cme4}\\
   \left\{ -\frac{\hbar^2}{2m}
   \frac{\partial^2}{\partial \rr^2} + V_R(\rr')
   + U_0 |\Phi(\rr)|^2\right\} \Phi(\rr) = \mu \Phi(\rr).
\label{gpe4}
\ena
We have introduced here the condensate wave function $\Phi=\sqrt{N}\chi$ and dropped the prime on the relative coordinate $\rr$. The effective potentials for the center-of-mass and relative motion are defined as 
\bea
   V_C(R) = \frac12 M\omega^2(1+4\lambda \frac{\langle r^2\rangle}{\aos^2}) R^2 + 
   \frac{\lambda M \omega^2}{2\aos^2} R^4,\label{cmpot}\\
   V_R(r) = \frac12 m\omega^2(1+4\lambda \frac{\langle R^2\rangle}{\aos^2}
   ) r^2 + \frac{\lambda m \omega^2}{2 \aos^2} r^4,
\label{gppot}
\ena
where $\langle R^2 \rangle = \int J d\RR R^2|\psi_C(\RR)|^2$ and $\langle r^2 \rangle = N^{-1}\int d\rr r^2|\Phi(\rr)|^2$. Averages of odd powers in the relative coordinate vanish because the relative coordinates are centered around the CM.
The quantities $E_C$ and $\mu$ are introduced as Lagrange multipliers to ensure separate 
normalization; $E_C$ is actually 
the center-of-mass energy and $\mu$ is the chemical potential as usual, and $M=Nm$ is the total mass of the gas. For harmonic trapping ($\lambda=0$), the equations decouple: the CM behaves like a single 
particle of mass $M$, and the condensate wave function for the internal motion obeys the Gross-Pitaevskii equation. 
The same effective potential for the internal motion was derived in Ref.\ \cite{zimmermann}; however, the Gross-Pitaevskii equation was there taken as a starting point.

In the absence of rotation, the spatial extent of the CM wave function is on the order of the effective CM oscillator length $a_C=(\hbar/M\omega)^{1/2}$ while
the size of the condensate is of order $\aos$, which is a factor 
$N^{1/2}$ larger. As a result, the CM motion has a negligible effect on 
the motion of the condensate, and the usual Gross-Pitaevskii equation is 
retained. However, we are interested in the case of highly excited CM motion, where the total angular momentum $L$ carried by the CM is comparable to the number of particles $N$, and 
in that limit, the effects of the CM wave function cannot be neglected.

Numerical solutions to the coupled equations of motion 
(\ref{cme4}-\ref{gppot}) are obtained with the steepest descent method. 
The $z$ dependence of the CM wave function $\psi_{C}$ decouples and is 
exactly Gaussian, while for the condensate wave function $\Phi$ both the 
radial and axial dependence is subject to numerical minimization. We have 
chosen the harmonic part of the potential to be isotropic, 
$\omega_z=\omega$.
For large values of $L$, the CM wave function becomes very peaked
and difficult to handle numerically.
It has therefore been necessary to restrict the number of particles $N$ to 
$N\leq 15$, and the possible angular momenta $L$ to a few times $N$. 

\begin{figure}
\includegraphics[width=\columnwidth]{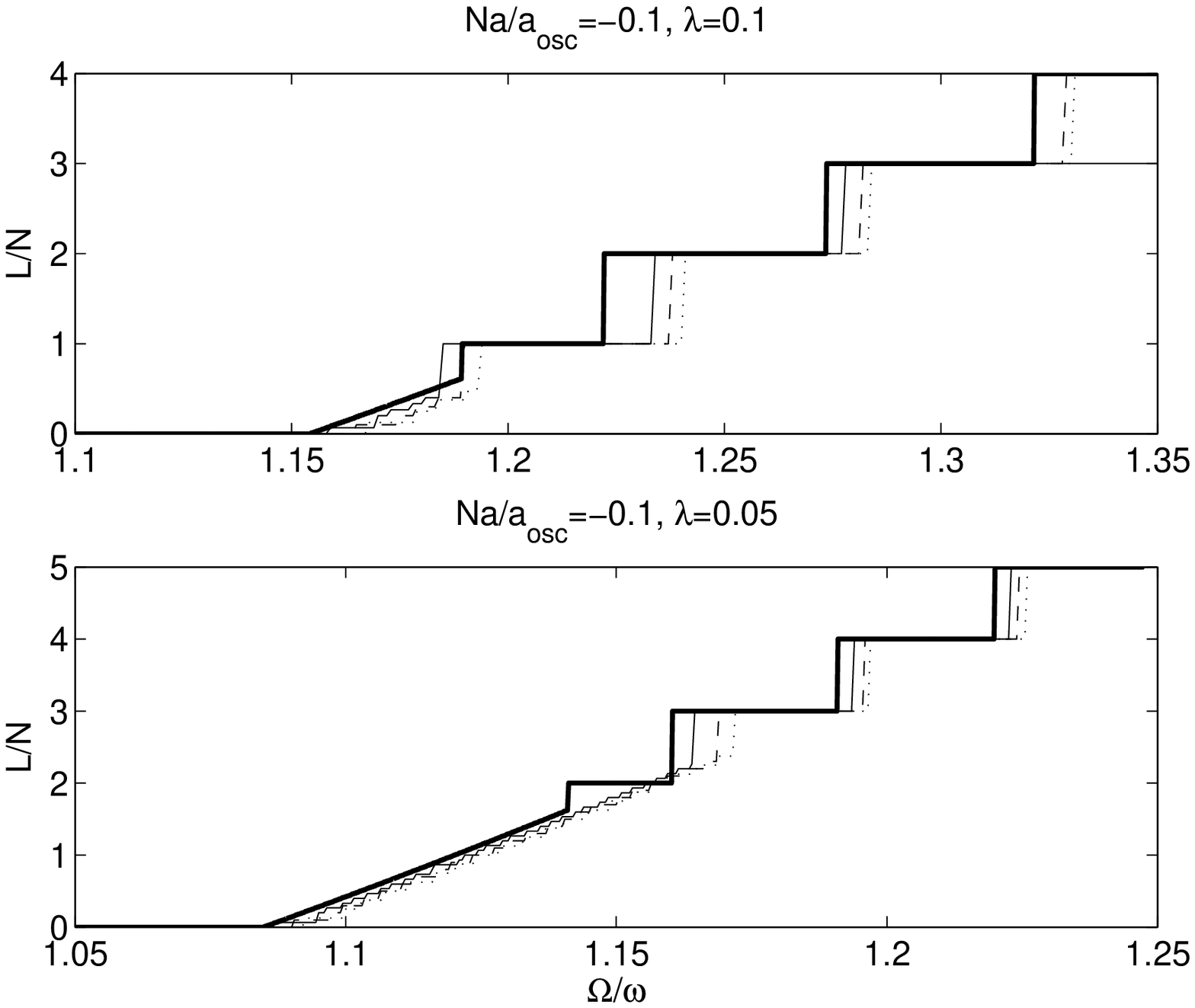}
\caption[]{Dependence of angular momentum on rotational 
frequency. The coupling $Na/\aos$ and anharmonicity $\lambda$ 
are as indicated above each panel. The thick solid lines indicate the 
variational 
result, and the thinner lines represent the fully numerical results for the 
cases $N=15$ (full), 10 (dashed) and 8 (dotted). When the angular momentum 
is carried by the center of mass, 
the curves display a smooth increase, but when vortices appear in the 
condensate, the angular momentum is fixed to integer values.
\label{fig:omegal}}
\end{figure}

The result for the total angular momentum is 
shown in Fig.\ \ref{fig:omegal}. We have chosen $Na/\aos=-0.1$, and 
$\lambda=0.05$ and $\lambda=0.1$ as two illustrative examples. Both these 
parameter values lie in the region where for slow rotation the CM carries 
the motion, hence the total angular momentum per particle $L_z/N$ increases 
only in steps of $\hbar N^{-1}$, but 
for faster rotation the angular momentum is carried by a vortex in the 
condensate wave function, and the angular momentum per particle is integer.
For sufficiently large $\lambda$ and weak attraction, vortex states are 
energetically favorable for all $\Omega$, because the anharmonic potential 
favors vortex states where the density distribution is toroidal. Conversely, for 
large enough $Na/\aos$ and small enough $\lambda$, the attraction between 
bosons favors the state of CM rotation where the density distribution is
maximally concentrated. 

The numerical results give at hand that a mixed state, where both $m_l$ 
and $L$ are nonzero, is never the energetically 
favorable situation. This is clear, since the density distribution of 
such a state is a toroid that is not centered at the middle of the 
trap, but occupies regions of high potential energy in the 
Mexican-hat shaped effective potential formed by the anharmonic and 
centrifugal energies. 

In order to be able to cover larger parameter spaces, we must consider a 
variational solution of the equations of motion 
(\ref{cme4}-\ref{gppot}).
We choose trial functions with Gaussian envelope for the CM and condensate wave functions:
\bea
   \psi_C(\RR) &=& {\mathcal N}_C (X+iY)^L 
   e^{-R^2/2ba_C^2-(\omega_z/\omega) Z^2/2a_{C}^2},
   \nonumber\\
   \Phi(\rr) &=& {\mathcal N}_R (x+iy)^{m_l} 
   e^{-r^2/2c\aos^2-(\omega_z/\omega)z^2/2(d\aos)^2}.
\ena
That this is a good ansatz has been confirmed by inspecting the shape of 
the numerically calculated wave functions.
${\mathcal N}_C$ and  ${\mathcal N}_R$ are normalization constants.
As already noted, the $z$ dependence of the CM wave function is not subject to variational minimization, but is a Gaussian with the fixed width $a_{C}\sqrt{\omega/\omega_z}$. On the other hand, the widths $b$, $c$ and $d$ are variational parameters. The ansatz allows for the angular momentum to be shared between the CM 
and internal motion and thus incorporates both the known limits of a BEC in a vortex state for $U_0=0$, and CM rotation for $\lambda=0$. The total angular momentum is $L_z=(L +Nm_l)\hbar$; 
we have already seen that the lowest-energy states in a rotating frame 
have either $L=0$ or $m_l=0$. Defining $q=L/N$, the energy is
\bea
   E = \frac{N}{2}q\hbar\omega \left( \frac{1}{b} + b+4(m_l+1)\lambda b c + 
   \lambda q b^2
   \right) \nonumber\\
   + \frac{N}{2}(m_l+1)\hbar\omega\left[\frac1c + c + \lambda (m_l+2)c^2\right]
   + \frac{N}{4}\hbar\omega_z\left(\frac1{d^2} + d^2\right) 
   \nonumber\\
   + N\hbar\omega  \left(\frac{\omega}{2\pi \omega_z}\right)^{1/2} 
   \frac{N a}{\aos}
   \frac{(2m_l)!}{(m_l!)^2 2^{2m_l}} \frac{1}{c d}.
\label{variationalcenergy}
\ena
Terms of order unity are discarded. 
The energy is now minimized with respect to the parameters 
$b$, $c$ and $d$, and the energetically optimal way of distributing 
the total angular momentum between the CM and internal motion is 
determined. The result is plotted as solid lines in Fig.\ \ref{fig:omegal}.
It should be noted that 
the variational solution is carried out for the limit of large $N$ while
the full numerical solution is restricted to finite values of $N$.

\begin{figure}
\includegraphics[width=\columnwidth]{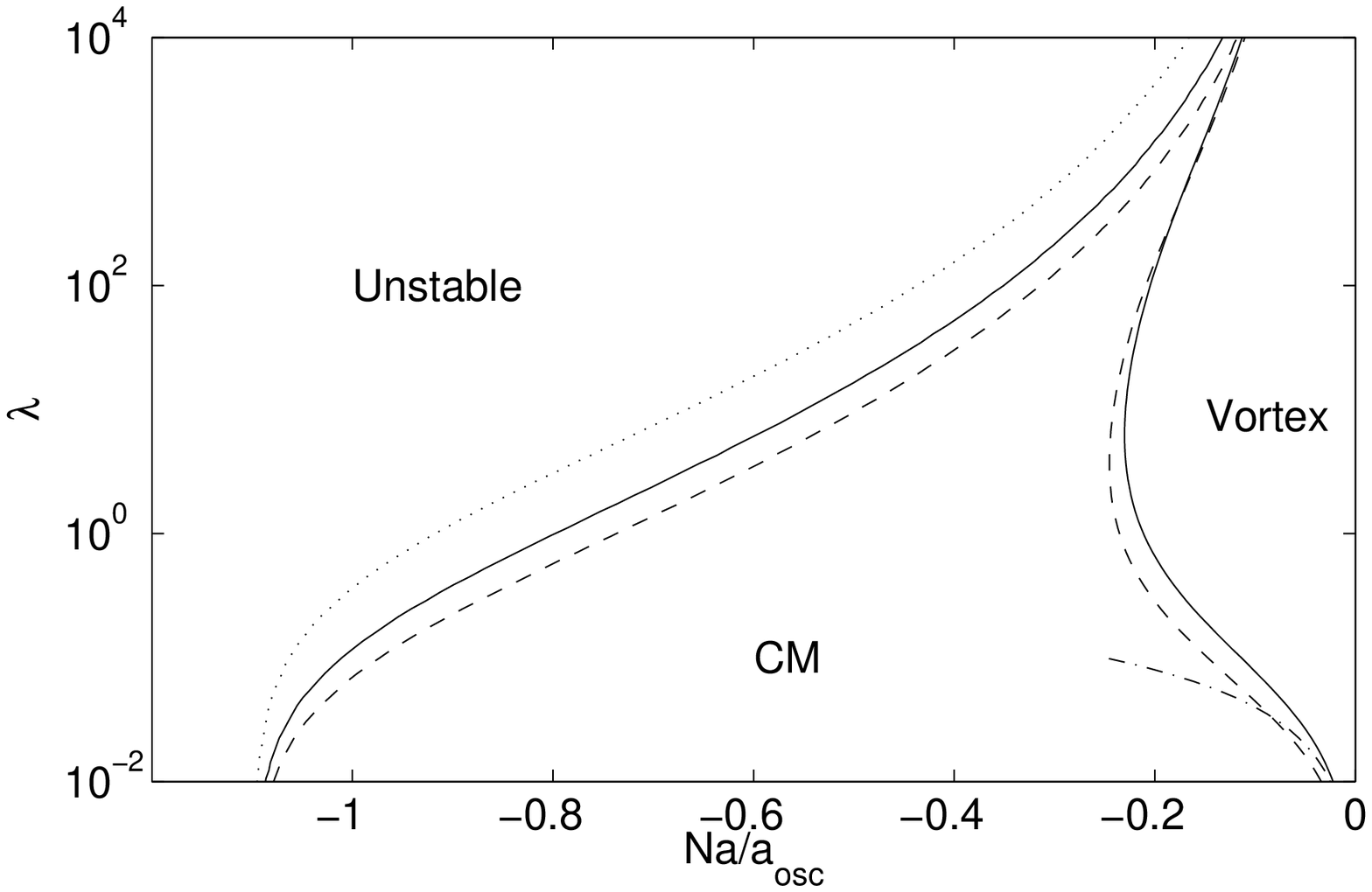}
\caption[]{The solid lines represent the boundaries 
between the three phases; vortex phase, center-of-mass rotation 
and unstable phase; for the case of one quantum of angular momentum per 
particle, $L_z=N\hbar$. 
The dashed lines represent the case $L_z=2N\hbar$. 
The dotted line is the boundary between the stable and unstable 
regimes for a nonrotating cloud. The lines are calculated in the 
variational scheme. The dot-dashed line is the perturbative result.
\label{fig:xover}}
\end{figure}

Fig.~\ref{fig:xover} shows the boundary between the CM-rotation and vortex regimes for a few fixed angular momenta as calculated in the variational scheme. As expected, a large $\lambda$, large angular momentum and weak 
coupling favors the vortex state. The coupling at which the cloud 
becomes unstable against collapse is also shown; the 
apparent shrinking of the stable regime with increasing $\lambda$ is 
due to the decrease of the effective size of the cloud, and therefore
an increase in density, when $\lambda$ is increased at fixed $\omega$.
Included is also the result of a perturbative calculation, treating the 
exact many-body wavefunctions to first 
order in $\lambda$ and $Na/\aos$ (cf. \cite{wilkin,mottelson}); it is seen 
that perturbation theory only works when these quantities are very small. 

In conclusion, a Bose gas 
with negative {\it s}-wave scattering length confined in an anharmonic trap 
will responds to a rotational force by either exciting center-of-mass 
rotational motion or forming a multiply quantized vortex, depending 
on the strength of the interaction and the anharmonic term. 
The wave function of the CM position and the condensate wave function associated with the
internal motion are described by two equations that in the general 
anharmonic case are coupled. These results show that the behavior of condensates with attractive interactions is markedly dependent on the harmonicity of the potential, and that an anharmonic trapping potential is expected to allow for vortex creation.
Reference \cite{dalibard} reported experiments using an anharmonic potential with a small quartic term ($\lambda \approx 0.001$), created by imposing a Gaussian optical potential over a magnetic trap. A similar setup, possibly with a stronger anharmonicity through increased laser power or decreased beam width, could be employed in order to explore the rotational states discussed in this paper.

While preparing this manuscript we became aware of the related work by Ghosh 
\cite{ghosh}.

Discussions with C.~J.\ Pethick are gratefully acknowledged.
The authors acknowledge support from the Academy of Finland 
(Project 50314), the European Network ``Cold Atoms and Ultra-Precise Atomic Clocks'' (CAUAC), and the Magnus Ehrnrooth foundation.

\begin{thebibliography}{99}
\bibitem{pethicksmith} C.\ J.\ Pethick and H.\ Smith, {\it Bose-Einstein 
Condensation in Dilute Gases} 
(Cambridge University Press, Cambridge, 2001).
\bibitem{wilkin} N.\ K.\ Wilkin, J.\ M.\ F.\ Gunn, and R.\ A.\ Smith, 
Phys.\ Rev.\  Lett.\ {\bf 80}, 2265 (1998).
\bibitem{nozieres} P.\ Nozi{\`e}res and D.\ Saint James, J.\ Phys.\ 
(Paris) {\bf 43}, 1133 (1982). 
\bibitem{hussein} M.~S.\ Hussein and O.~K.\ Vorov, Phys. Rev. A {\bf 65}, 
053608 (2002).
\bibitem{pethickpitaevskii} C.\ J.\ Pethick and L.\ P.\ Pitaevskii, 
Phys.\ Rev.\ A {\bf 62}, 033609 (2000).
\bibitem{mottelson} B.\ Mottelson, Phys.\ Rev.\ Lett.\ {\bf 83}, 
2695 (1999).
\bibitem{fetterrokhsar} A.\ L.\ Fetter and D.\ Rokhsar, Phys.\ Rev.\ 
A {\bf 57}, 1191 (1998).
\bibitem{kohn} W.\ Kohn, Phys.\ Rev.\ {\bf 123}, 1242 (1961).
\bibitem{elliottskyrme} J.\ P.\ Elliott and T.\ H.\ R.\ Skyrme,
Proc.\ R.\ Soc.\ London Ser.~A {\bf 232}, 561 (1955). 
\bibitem{lundh2002} E.~Lundh, Phys.\ Rev.\ A {\bf 65}, 043604 (2002).
\bibitem{leggett} A.\ J.\ Leggett, Rev.\ Mod.\ Phys.\ {\bf 73}, 307 (2001).
\bibitem{zimmermann} H.\ Ott, J.\ Fort\'agh, and C.\ Zimmermann, 
J.\ Phys.\ B {\bf 36}, 2817 (2003).
\bibitem{dalibard} V.\ Bretin, S.\ Stock, Y.\ Seurin, and J.\ Dalibard, 
cond-mat/0307464.
\bibitem{ghosh} T.\ K.\ Ghosh, cond-mat/0306563.
\end{thebibliography}
\end{document}